\documentstyle[manuscript,aps]{revtex}
\begin{document}
\begin{center}
{\LARGE Chaotic wave functions and exponential convergence of low-lying
energy eigenvalues}\\[0.5cm]

{\bf Mihai Horoi$^{1}$, Alexander Volya$^{2}$ and Vladimir
Zelevinsky$^{2}$}\\[0.2cm]

{\sl $^{1}$Physics Department, Central Michigan University, Mount Pleasant, MI
48859, USA\\
$^{2}$Department of Physics and Astronomy and\\
National Superconducting Cyclotron Laboratory,\\
Michigan State University, East Lansing, MI 48824-1321, USA}
\end{center}

\begin{abstract}
We suggest that low-lying eigenvalues of realistic quantum many-body
hamiltonians, given, as in the nuclear shell model, by large matrices,
can be calculated, instead of the full diagonalization, by the
diagonalization of small truncated matrices with the exponential extrapolation
of the results. We show numerical data confirming this conjecture. We argue
that the exponential convergence in an appropriate basis may be a generic 
feature of complicated (``chaotic") systems where the wave functions are 
localized in this basis. 
\end{abstract}

\newpage

Statistical properties of complex quantum systems have been studied extensively
from various viewpoints. The seminal papers by Wigner \cite{wig} and Dyson
\cite{Dyson} developed the random matrix theory (RMT \cite{Mehta})
where the systems are
considered as members of a statistical ensemble, and all hamiltonians of given
global symmetry appear with certain probabilities. The canonical Gaussian 
ensembles \cite{Brody,Haake} 
correspond to systems with very complicated dynamics 
when, in almost all bases connected by the transformations preserving
global symmetry, the components of generic eigenfunctions 
are uniformly distributed on the unit sphere in multidimensional Hilbert space.
On local scale, Gaussian ensembles predict specific correlations and
fluctuations of spectral properties which are in agreement
with empirical data for atoms, nuclei \cite{porter}, quantum dots \cite{markus}
and resonators (microwave \cite{cavity} and acoustic \cite{acou} experiments).
These spectral features are considered usually as signatures of quantum chaos
\cite{Haake,BW,ann}.

Recently, the detail studies of highly excited states in realistic atomic
\cite{grib} and nuclear \cite{big} calculations demonstrated that such
many-body systems are close to the RMT limit although they reveal some
deviations, partly due to the presence of the mean field \cite{MF}, 
coherent components \cite{LBBZ} of the residual interaction and its
two-body character \cite{Brody,Fla}. 
In complex atoms and nuclei, precise experimental
information exists, as a rule, about low-lying states only. Effective residual
interactions , such as the
Wildenthal-Brown interaction \cite{WB} for the $sd$-shell model 
turned out to be successful well beyond the input used for their original 
fit. This justifies the use of such interactions for studying generic
complicated states in the region of high level density. The whole 
shell model approach is based on the large-scale diagonalization even if one 
is interested in the low-lying states only. The dimensions of matrices increase
dramatically with the number of valence nucleons 
which makes the full diagonalization impractical, even after projecting out 
correct angular momentum and isospin states. This problem is avoided in the
Monte Carlo shell model method \cite{MC}, but, apart from the so-called
sign problem \cite{YA} which requires the introduction of an extrapolation when
working with realistic interactions, this method is better suited for
calculating thermal properties or strength functions than 
spectroscopic characteristics. In order to consider individual
levels, one needs to supplement the Monte Carlo sampling with some variational
procedure including an additional ``stochastic" 
diagonalization (\cite{Frick} or the QMCD approach \cite{otsu}).
An important step towards larger dimensions in the standard shell model
is made with the development of the DUPSM code \cite{akiva}.

Here we suggest a simple approach which allows one, for calculating
energies of low lying states, to reduce large dimensions
of matrices under study by orders of magnitude, while keeping high
precision of the results. The approach is based on the statistical properties
of complicated many-body states \cite{grib,big}. Because of the strong residual
interaction and ``geometric chaoticity" \cite{big} of the angular momentum
coupling, the eigenstates are extremely complex superpositions of
independent particle Slater determinants. However, in contrast to
the limiting case of the Gaussian orthogonal ensemble (GOE), the stationary wave
functions are not fully delocalized in shell model space. Due to
inherently self-consistent nature of the residual interaction (even if it is
extracted in a semiempirical manner), its strength does not exceed the typical
spacings between single-particle levels which are determined by the mean field,
i.e. by the same original forces. Together with the fact that the two-body
forces cannot couple very distinct configurations, this leads to a
band-like structure of the hamiltonian matrices in the shell model basis.

Theory of banded random matrices did not reach the same degree of
completeness as that of canonical Gaussian ensembles. Nevertheless, both
mathematical \cite{Fyod} and numerical \cite{Izr} arguments favor the
localization of the eigenstates in Hilbert space, similar to the coordinate
localization of electronic states in disordered solids. The generic many-body
states in complex atoms or nuclei have a typical localization width
\cite{big,Izr,entr}.
Inversely, the simple shell model configurations are packets of the
eigenstates. Their strength function is fragmented over the range of energies
characterized by the spreading width $\Gamma$ which is nearly constant along
the spectrum because the coupling matrix elements between the complicated
states are small just as it is needed to compensate small level spacings in the
region of high level density \cite{SF,HRW,Wies}. The qualitative arguments are
confirmed by more general theory \cite{LBBZ} as well as by detailed 
numerical calculations for atoms \cite{grib} and nuclei \cite{big,fraz}.
The nuclear case is close to the strong coupling limit \cite{bert,lew,big}
where the typical width can be estimated \cite{fraz} as $\Gamma\approx 2
\bar{\sigma}$ in terms of the energy dispersion of a simple configuration
$|k\rangle$,
\begin{equation}
\sigma_{k}^{2}=\langle k|(H-\langle k|H|k\rangle)^{2}|k\rangle =\sum_{l\neq k}
|H_{kl}|^{2}.                                                  \label{1}
\end{equation}
Here $H_{kl}$ are the off-diagonal elements of the residual interaction between
the basis states so that the calculation of (\ref{1}) does not require any
diagonalization. The dispersions $\sigma_{k}$ of different simple states
fluctuate weakly \cite{big} and in our estimate of $\Gamma$ they are
substituted by the appropriate mean value $\bar{\sigma}$ which can be found by
the methods of statistical spectroscopy \cite{FR}.

The practical method of truncating large shell model matrices
was suggested in \cite{trunc}. The shell model states are 
grouped into partitions (sets of states belonging to the same particle 
configuration). Since the states separated 
in energy by an interval broader than $\Gamma$
are not significantly mixed with the studied state, we 
truncated the matrix retaining only the partitions whose statistical centroids 
$\bar{E}=\overline{\langle k|H|k\rangle}$ are closer than 3$\sigma$. The 
spin-isospin projection and the elimination of the center-of-mass admixtures 
can be done within the truncated subspace only. In order to keep the correct 
shell model structure, the partitions should be included as a whole. As shown 
in \cite{trunc}, this method allows for the calculation of low-lying 
energies with sufficient precision in large shell-model spaces. The 
truncated eigenvectors overlap
with the exact ones on the level of better than 90\%.

Going beyond the simple truncation, we consider the convergence of 
level energies to the exact values as a function of the increasing dimension 
$n$ of the diagonalized matrix. As an example we take the $^{51}$Sc nucleus
where the $pf$-shell model dimensions of $1/2^{-}$ and 
$3/2^{-}$ states are 13016 and 24474, respectively. Spectroscopic information
 on this
radioactive isotope is only tentative providing an interesting experimental and
theoretical problem. For the calculations, the FPD6 interaction \cite{rich}
was used. Fig. 1 shows the calculated energies of the two lowest 
$3/2^{-}$ states and two lowest $1/2^{-}$ states 
for several values of $n$ ranging from $n=2000$ to the full dimension $N$.
Already the smallest dimensions lead to a good agreement within few
As the dimension
increases, in all four cases the running eigenvalue
converges very fast and monotonously to the exact result. The convergence is 
almost pure exponential, $E(n)=E_{\infty}+A\exp(-\gamma n)$;
typically $A\approx 300$keV, $\gamma\approx 6/N$.

The exponential convergence of eigenvalues would be extremely helpful for 
shell model practitioners. It would make almost redundant the full large scale
diagonalization if one is interested in the low-lying states only. Instead, the
calculations for
several increasing dimensions (still far from the full value and therefore
easily tractable) would end in
determining the exponential parameters and simple extrapolation to the exact
result. At present, the rigorous mathematical theory of convergence is absent,
and we limit ourselves by qualitative arguments and plausible conjectures.

The convergence under consecutive truncations is determined by the type of the 
matrix and by the original unperturbed basis which orders the basis vectors in 
a certain way. The ordering 
is done almost uniquely in the spherical shell model where
the mean field is fixed and all many-body states are organized in partitions.
For the lowest levels, the admixtures of highly excited states outside of the
starting truncation correspond to the wings of the strength function. As
confirmed by the atomic \cite{grib} and nuclear \cite{big,fraz} studies, the
strength function has in average universal shape. This shape evolves from the 
standard Breit-Wigner function \cite{BM} for the ``weak damping" case to the 
Gaussian form at strong
damping (semicircle in the RMT limit of the uniform spectra 
\cite{LBBZ,Fyod} which is not reached in practical cases). Correspondingly,
the dependence of the spreading width on the strength of the residual
interaction changes from quadratic in the standard golden rule \cite{BM} to
linear \cite{lew,LBBZ,fraz}. The remote wings 
of the strength function have their own energy behavior \cite{grib,fraz}.
With high accuracy they can be described \cite{fraz} by
an exponential function of energy. This is a clear manifestation of the
localization of the eigenfunctions typical for the banded hamiltonian
structures \cite{cas,Fyod}. In this limiting regime, the average local
strength is simply proportional to the total remaining strength,
$F(E)\approx {\rm const}\,[1-\int ^{E}dE'\,F(E')]$ which gives $F(E)\sim
\exp(-{\rm const}E)$. The exponentially weak mixing should lead to the
exponentially small energy shifts and to the corresponding convergence of the
eigenvalues.

We can expect the exponential behavior to be generic for the large matrices of
quasi-banded form with the off-diagonal elements of approximately the same
order of magnitude along the spectrum, This conjecture can be checked by
generating random matrices with the desired properties and diagonalizing them
in a sequence of progressing truncations (the matrices are first ordered
according to their diagonal elements). Since there is no ``vertical"
structure in such random matrices, we do not have here physical arguments
concerning the optimal truncation sequence. We show in Fig. 2, left, a typical
result for the banded GOE matrix with the width $b=0.293 N$
which clearly demonstrates the exponential
convergence. The full GOE matrix, Fig. 2, right, converges more
slowly in the absolute sense and does not saturate. This is a simple
consequence of the fact that the ground state of a GOE matrix is repelled by
the higher states to the edge of the semicircle (-2 with the GOE definition
accepted here and in \cite{big}). This process is driven
by the off-diagonal elements; all of them in average have 
the same order of magnitude. Their number and, whence, the dispersion
$\sigma$, eq. (\ref{1}), is greater in the full GOE. Therefore the distance
from the unperturbed position is also greater in this case (the diagonal and
off-diagonal contributions add in quadratures). In all studied cases with the
width $b$ changing from $0.1 N $ to the full GOE, the convergence is
exponential and the exponent $\gamma$ is approximately scaled inversely
proportional to $b$. 

In realistic cases there is also a leading sequence of regular diagonal 
elements. A similar banded matrix example, considered in \cite{grib}, 
goes back to Wigner 
\cite{wig}. The matrix consists of the equidistant diagonal
with the spacing $D$ and random off-diagonal matrix elements $V_{kl}$ within
the band $|k-l|\leq b$. At relatively weak interaction, $g\equiv
\langle V^{2}\rangle/D^{2}<1$, the main contribution 
in the perturbation series for the
admixture $w_{n}=C_{n}^{2}$ of a very remote state $|n\rangle, \, n\gg 1$,
to the wave
function of a low-lying state $|0\rangle$ is given by the summation of 
long ``straight" paths in Hilbert space connecting the initial state with the
final one through various intermediate stops. 
Because of the random character of the off-diagonal interaction, the
mean value of $w_{n}$ is determined by the squares of the contributions of
these paths (no interference). In the approximation of a weakly changing level
density, this can be approximately
written as an integral equation
of the random-walk type,
\begin{equation}
w_{n}=\frac{g}{n^{2}} f_n+\frac{g}{n^{2}}\sum_{k}f_{n-k}w_{k}
                                                           \label{10}
\end{equation}
where the factor $1/n^{2}$ comes from the energy denominators, and $f_{n}$
shows the behavior of typical squared
off-diagonal matrix elements $V_{n0}^{2}$ as a function of the
distance $n$ from the diagonal. With the sharp band boundary \cite{grib}, 
the weights $w_{n}$ decrease very fast, essentially as $\exp(-n\ln n)\sim
(n!)^{-1}$. With the smooth cut-off, the convergence is getting closer to 
exponential. Thus, for the exponential cut-off of the matrix elements, 
$V_{kl}\sim\exp(-|k-l|/b)$, we have $f_{n}=\exp(-2n/b)$, and 
eq. (\ref{10}) allows a simple solution $w_{n}=A\exp(-2n/b)/n^{2}$. 
Therefore the
contributions to energy of the state $|0\rangle$ should converge 
$\sim nDw_{n}=DA\exp(-2n/b)/n$. Fig. 3 illustrates this consideration
by an example of the numerical diagonalization of a random matrix with 
the equidistant diagonal and the exponential cut-off. One may note that the
rate of convergence is similar to that in the shell model calculation, see 
above, where the effective width of the band is close to $b\approx N/4$
\cite{big}. However, the
method of eq. (\ref{10})
becomes invalid in the case of strong interaction when the contributions
to the perturbation series of additional loops in Hilbert space cannot be
neglected. The range of convergence is seen from the expression for the
constant in the above solution, $A=g/[1-g\sum_{k=1}k^{-2}]$, which determines
the critical value $g_{c}=6/\pi^{2}$.

It is interesting to test the character of the convergence in simple 
solvable models. 
A harmonic oscillator, shifted from the 
equilibrium position by a constant force, $H=a^{\dagger}a+
\lambda(a+a^{\dagger})$, 
lowers its energy by $\lambda^{2}$. The exact ground
state is a coherent combination of unperturbed states $|n\rangle$. 
In accordance with the composition of the coherent state, the convergence of
the energies in the unperturbed basis of the original oscillator is  
$\sim \lambda^{2n}/n!$. This is clearly seen in Fig. 4, left. 
The fast convergence
is due to the constant level density along the main diagonal while the
perturbation has matrix elements growing only $\sim n^{1/2}$. In the case of a
quartic anharmonic oscillator, the exponential convergence is modulated with
oscillations.

Another example displays the case of the 
slow convergence. The tight-binding
model of a finite one-dimensional lattice has degenerate levels in each of $N$
identical wells and the amplitude $v$ of hopping between the adjacent wells.
The eigenstates of the model are delocalized standing Bloch waves with energies
within the band, $E_{q}=2v\cos\varphi_{q}, \; \varphi_{q}=\pi q/(N+1), \, 
q=1,2,\dots,N$. As it is easy to see, the truncation in the site basis 
corresponds to the convergence $\sim 1/n^{2}$, see Fig. 4, right. 

Tridiagonal
matrices with the entries $H_{nn}\equiv \epsilon_{n}$ and $H_{n-1,n}=H_{n,n-1}
\equiv V_{n}$ smoothly depending on $n$ can be analyzed in a general way using
the recurrence relation for the secular determinants $D_{n}(E)$ of the matrix
$(H-E)$ truncated at the $n$th step,
\begin{equation}
D_{n}(E)=(\epsilon_{n}-E)D_{n-1}(E)-V_{n}^{2}D_{n-2}(E).          \label{11}
\end{equation}
The consecutive approximations $E^{(n-1)}$ and $E^{(n)}, \; n\gg 1,$ to a 
low-lying eigenvalue $E$ are the roots of $D_{n-1}(E^{(n-1)})=0$ and
$D_{n}(E^{(n)})=0$, respectively. For $\epsilon_{n}\gg E^{(n)}$ (with a slight
modification, the method works also for initially degenerate matrices with
$\epsilon_{n}=$const), the asymptotic
behavior of the energy increments $\Delta_{n}=E^{(n)}-E^{(n-1)}$ follows from
(\ref{11}) as 
\begin{equation}
\frac{\Delta_{n}\Delta_{n-2}}{(\Delta_{n}+\Delta_{n-1})(\Delta_{n-1}+
\Delta_{n-2})}=\frac{V_{n}^{2}}{\epsilon_{n}\epsilon_{n-1}}\equiv\lambda_{n}.
                                                             \label{12}
\end{equation}
The exponential convergence corresponds to $\lambda_{n}\rightarrow\lambda=$ 
const$\neq 0$ at
large $n$ (similar increase of diagonal and off-diagonal elements). Then the
increment ratio $\xi_{n}=\Delta_{n}/\Delta_{n-1}$ also goes to a constant limit
$\xi=(1/2\lambda)[1-(1-4\lambda^{2})^{1/2}]$ which restricts the exponential
convergence region to $\lambda^{2}<1/4$. The existence of the constant limit 
is still compatible with an additional preexponential factor weakly dependent 
on $n$; at large $n$ corresponding fits are usually indistinguishable.
An explicitly solvable (by the
Bogoliubov transformation) model of the harmonic oscillator with the
perturbation $\lambda(a^{2}+a^{\dagger 2})$ agrees completely with this
estimate. The case $\lambda^{2}=1/4$ corresponds here to the degeneracy of the
oscillator with zero frequency, 
and at $\lambda^{2}>1/4$ the spectrum is inverted. In general, 
$\lambda=1/4$ resembles a critical point; the convergence here is described by
a power law being exponential outside of this region.

The main difference of the tight-binding case  
from the oscillator model is the degeneracy of the unperturbed levels
(absence of the leading diagonal) which results in the delocalized wave
functions of eigenstates. Because of the degeneracy, the analog of eq. 
(\ref{12}) contains, instead of $\lambda_{n}^{2}$, the ratio 
$v^{2}/E^{(n)}E^{(n-1)}$ which is just equal to the critical value 
1/4 in the limit of large $n$.
The situation is similar for the spin chains with the
nearest neighbor interaction where the finite size effects on the
ground state energy were repeatedly studied \cite{Woy} and corrections also go
as $n^{-2}$. We expect the presence of disorder
(random positions of the original site levels), which leads to the 
localization of the eigenfunctions, to be accompanied by the transition to
the exponential convergence of the eigenvalues.

In conclusion, we discussed the convergence of the low-lying eigenvalues of
large matrices describing the realistic many-body hamiltonians of shell-model
type. We gave arguments in favor of the conjecture that the exact
diagonalization of relatively small matrices, truncated 
on the grounds of physical partitions and
generic spreading widths of simple configurations, provides a starting
approximation which can be extrapolated to the exact result with the aid
of a simple exponential continuation. The arguments are based on the generic
features of quantum chaotic many-body dynamics, simple models 
and the results of numerical analysis.  \\
\\
The authors acknowledge support from the NSF grants 95-12831 and 96-05207.
They thank F.M.Izrailev for stimulating discussions.

\newpage
\begin{center}
FIGURE CAPTIONS
\end{center}
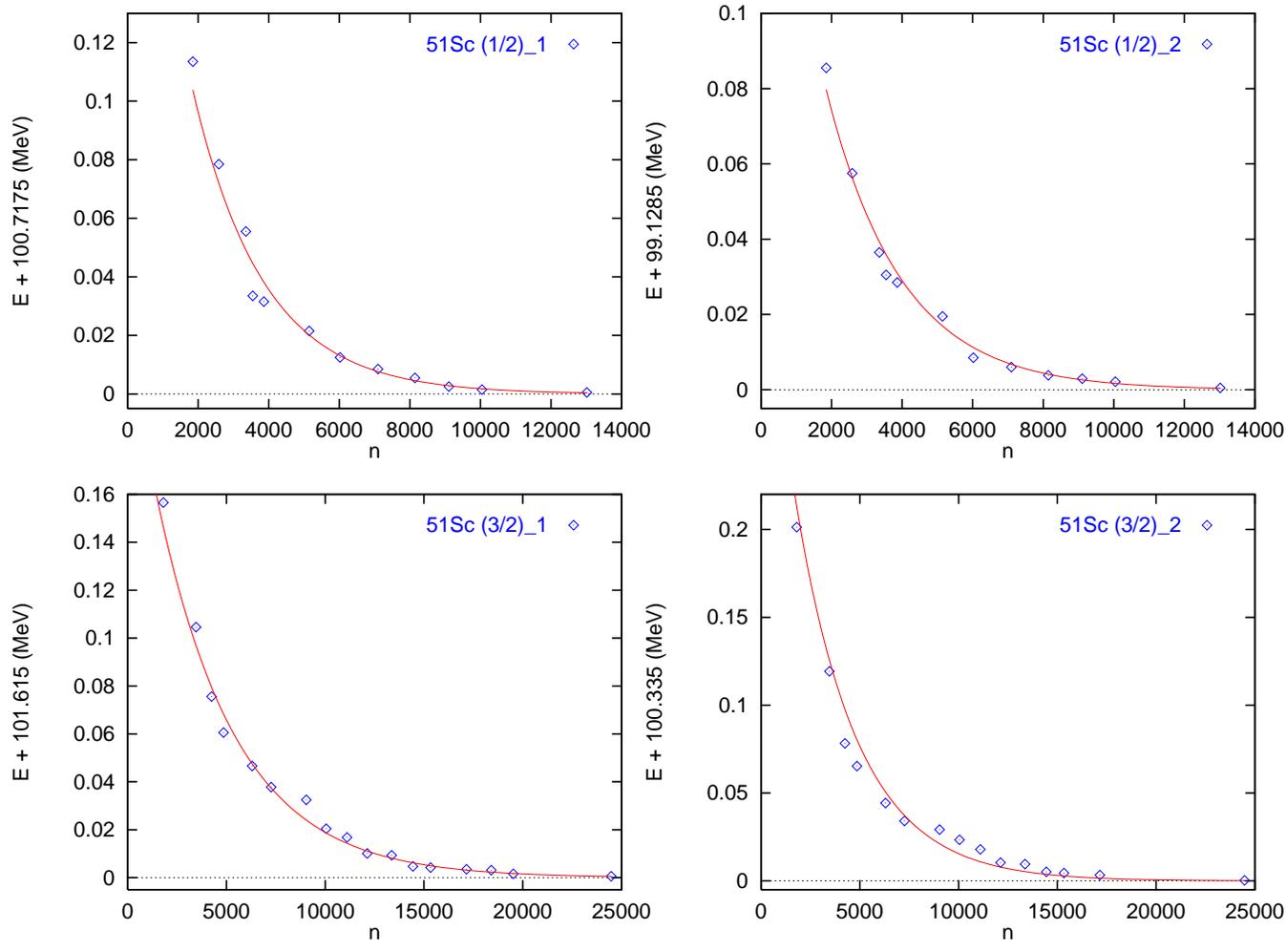
{\bf Figure 1}. Energy deviations from the exact shell model result for 
the lowest excited states $1/2^{-}$ and
$3/2^{-}$ in $^{51}$Sc (diamonds)
calculated as a function of the
progressive matrix truncation $n$. Solid lines give a fit $A\exp(-\gamma n)$
with the parameter values $A=0.26, 0.19, 0.23, 0.35$ MeV and $\gamma=49.7\times
10^{-5}, 47\times 10^{-5}, 25\times 10^{-5}, 30\times 10^{-5}$ for $1/2^{-}_{1},
1/2^{-}_{2}, 3/2^{-}_{1}$, and $3/2^{-}_{2}$, respectively.

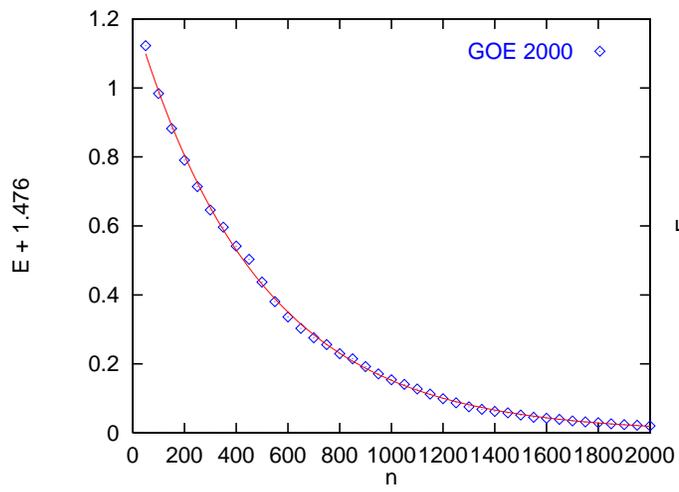
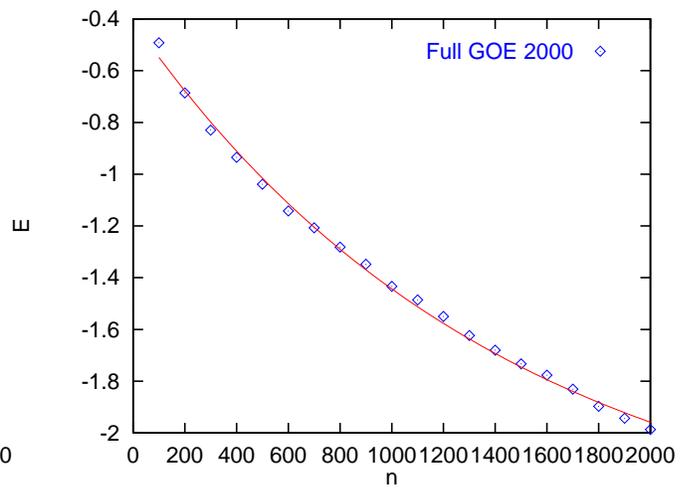
{\bf Figure 2}. Energy deviations for the ground state of random matrices
of dimension 1000 as a function of the progressive matrix
truncation $n$ (diamonds): the GOE-like banded matrix of the width $b=0.293 N$,
chosen in such a way that a half of the matrix elements vanish,
approximately in the same proportion as in typical shell model cases, left
panel; the full GOE matrix, right panel. Solid lines show
a fit $A\exp(-\gamma n)$ with
$A=1.18$ and $\gamma=3.8\times 10^{-3}$, left, and $A=1.96, \,\gamma=1.53\times
10^{-3}$, right. Note the absence of the horizontal asymptotics in the case of
the full matrix.

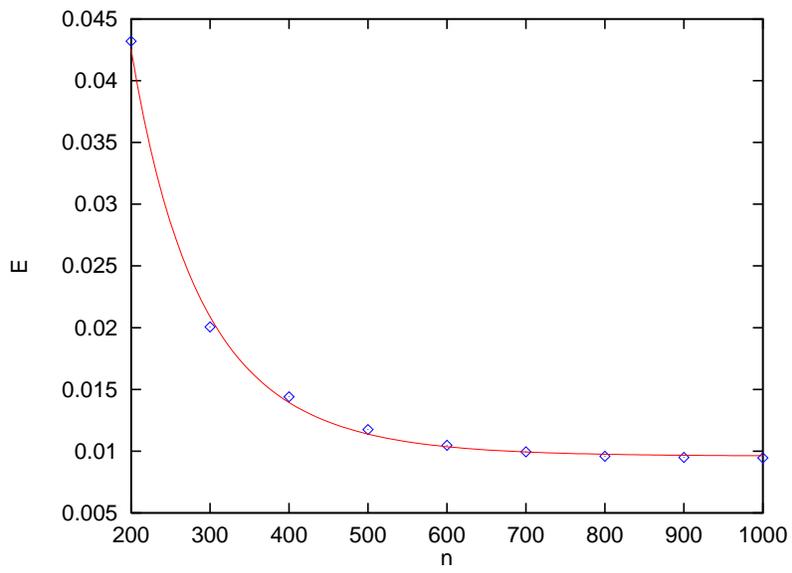
{\bf Figure 3}. Convergence of the ground state energy for the banded random
matrix with the exponential cut-off of matrix elements $V_{kl}\sim \exp(-|k-l|
/b)$ with $b=0.293 N$ (diamonds); the solid line is given by 
const$\,\exp(-2n/b)/n$ which corresponds to the solution of eq. (2).

{\bf Figure 4}. Convergence of the ground state energy in the tight-binding
model of a finite one-dimensional lattice with $\lambda$ as a hopping parameter,
left panels, and for a shifted harmonic oscillator with the hamiltonian
$H=a^{\dagger}a+\lambda(a+a^{\dagger})$, right panels. The upper parts show the
energy deviation $\Delta E_{n}=E_{0}(n)-E_{0}(\infty)$ as a function of the 
truncated dimension $n$ (solid lines for $\lambda =1$ and $\lambda=2$); dotted
lines show the predicted analytical convergence of the models, $\lambda\pi^{2}
/n^{2}$ (left) and $\lambda^{2n}/n!$ (right). The lower parts characterize the
convergence rate $\lambda_{n}\rightarrow \lambda$ 
by plotting $\lambda_{n}\equiv \Delta E_{n}n^{2}/\pi^{2}$, left, and 
$\lambda_{n}=(\Delta E_{n}n!)^{1/2n}$, right.

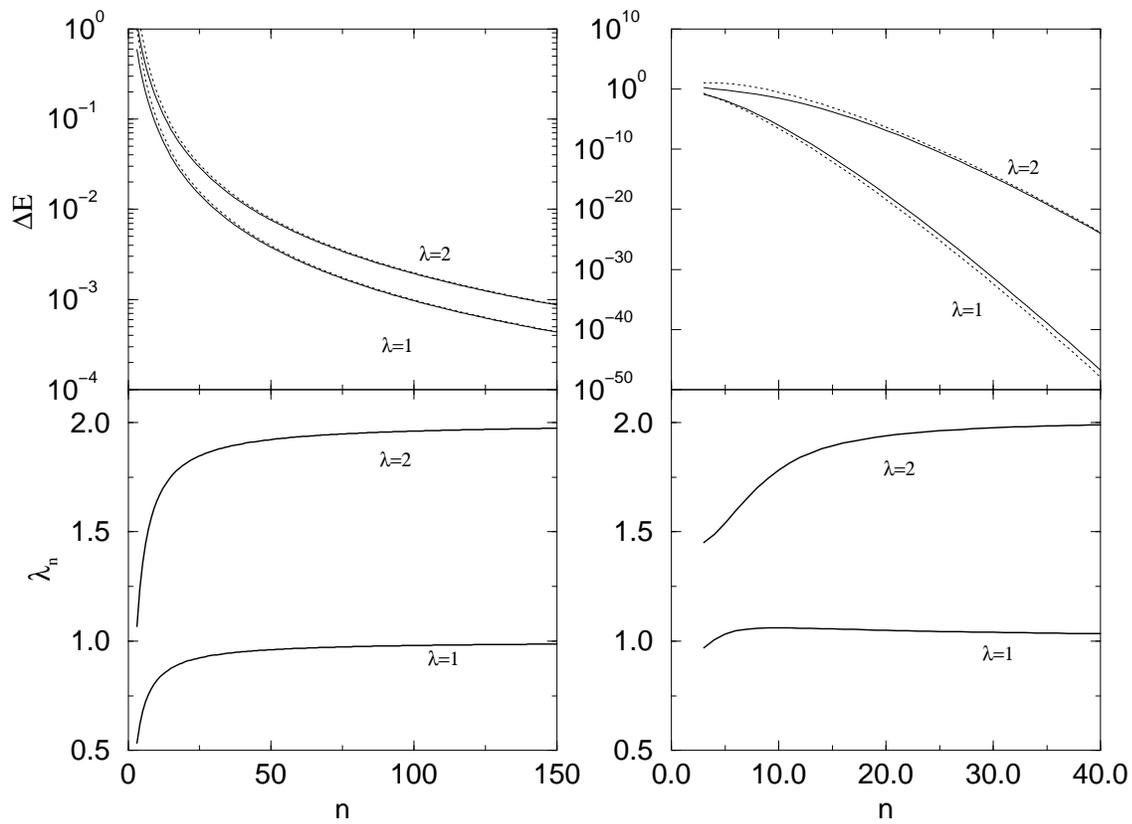
\end{document}